\documentclass[12pt, a4paper]{article}
\pdfoutput=1

\usepackage{amsmath}
\usepackage{amsfonts}
\usepackage{amssymb}
\usepackage{graphicx, rotating}
\usepackage{epsfig}
\usepackage{array}
\usepackage{latexsym}
\usepackage{graphicx}
\usepackage{xcolor}
\usepackage{amsmath,bm,amssymb}
\usepackage{cite}
\usepackage{slashed}
\usepackage{hyperref}
\usepackage{placeins}
\usepackage[normalem]{ulem}
\usepackage[utf8]{inputenc}
\usepackage[export]{adjustbox}
\usepackage{multirow, verbatim}

\newcommand{\be}{\begin{equation}}
\newcommand{\ee}{\end{equation}}
\newcommand{\bea}{\begin{eqnarray}}
\newcommand{\eea}{\end{eqnarray}}
\newcommand{\nn}{\nonumber}

\newcommand{\aNEW}{{\alpha_{\bar \theta}}}

\newcommand{\Tcr}{T_{\rm cr}}

\usepackage{mathtools}

\begin{document}


\begin{titlepage}

\begin{flushright}
\small
DESY 20-173\\
CERN-TH-2020-170
\end{flushright}
\vspace{.3in}

\begin{center}
{\Large\bf
Model-independent energy budget for LISA} \\ 
\vskip 0.2 cm
\bigskip\color{black}
\vspace{1cm}{
  {\large
Felix~Giese$^1$,
Thomas Konstandin$^1$,\\
Kai Schmitz$^2$,
Jorinde~van~de~Vis$^1$
}}
\vskip 0.3cm

{\small
$^1$DESY, Notkestra{\ss}e 85, D-22607 Hamburg, Germany\\
$^2$Theoretical Physics Department, CERN, 1211 Geneva 23, Switzerland,

}

\bigskip

\begin{abstract}
We provide an easy method to obtain the kinetic energy fraction in gravitational waves, generated during a cosmological first-order phase transition, as a function of only the wall velocity and quantities that can be determined from the particle physics model at the nucleation temperature. This generalizes recent work that achieved this goal for detonations. Here we present the corresponding results for deflagrations and hybrids. Unlike for detonations, the sound speed in the symmetric
phase also enters the analysis. We perform a detailed comparison between our model-independent approach and other approaches in the literature. We provide a Python code snippet to determine the kinetic energy fraction $K$ as a function of the wall velocity, the two speeds of sound and the strength parameter of the phase transition. We also assess how realistic sizable deviations in speed of sound are close to the phase transition temperature in a specific model.

\end{abstract}

\end{center}

\end{titlepage}

\newpage
\tableofcontents
\section{Introduction}
\label{sec:Introduction}

Gravitational waves (GWs) from cosmological phase transitions provide the intriguing possibility to probe high energy physics with upcoming interferometer experiments like the Laser Interferometer Space Antenna (LISA)~\cite{Audley:2017drz, Baker:2019nia}. If the phase transition is first order, it proceeds through the formation of bubbles. The expansion and collision of these bubbles can source a stochastic gravitational wave background~\cite{Maggiore:1999vm, Weir:2017wfa, Caprini:2018mtu, Christensen:2018iqi}. For phase transitions that are not too strongly supercooled, the main contribution to the gravitational wave signal is caused by sound waves in the plasma~\cite{Hindmarsh:2013xza, Hindmarsh:2015qta, Caprini:2019egz}. 

One important aspect to quantify the spectrum of this GW source is the energy budget of the phase transition. Usually, the fraction of kinetic energy that is available for gravitational wave production, $K$, is hereby determined from some phase transition strength parameter and the wall velocity using the bag equation of state to model the fluid filling the Universe~\cite{Espinosa:2010hh} for a single bubble expanding in a spherical way. That a single spherical bubble characterizes the energy budget well in all cases is an assumption that was called into question in recent simplified simulations~\cite{Jinno:2020eqg}. We will take this assumption for granted here, even though this point requires further scrutiny.

In a recent work~\cite{Giese:2020rtr}, we investigated what is the optimal phase transition strength parameter and all the relevant parameters for this analysis. For the case of phase transitions that lead to a detonation, we found that the optimal strength parameter is the quantity
\be
	\aNEW \equiv \frac{D\bar \theta(T_+)}{3w_+}  \, , \quad 
	 {\rm with} \quad \bar \theta \equiv e - \frac{p}{c_{s,b}^2} \, , 
\,
\ee
where $e$ is the energy density, $p$ the pressure, $w=p+e$ is the enthalpy density, and $c_{s,b}$ the speed of sound in the broken phase. The subscript $+$ denotes that the quantities are evaluated in front of the bubble wall and the $D$ indicates that the difference between the broken phase and the symmetric phase is evaluated. Moreover, the speed of sound in the broken phase enters the dynamics of the fluid such that, ultimately, the efficiency factor $\kappa$ can be determined using the strength parameter $\aNEW$ and the speed of sound $c_{s,b}$ alone. This approach is highly model independent: deviations from the full numerical analysis stem only from the temperature dependence of 
the speed of sound and deviations are typically below percent level. 

In the present work, we generalize this approach to deflagrations and hybrid solutions (Section~\ref{sec:hydro}). In addition to the above ingredients, also the speed of sound in the symmetric phase enters the analysis. This is unavoidable, since this speed of sound enters in the dynamics of the shock wave. At the same time, it is surprising that the kinetic energy fraction of hybrid solutions can be determined with high accuracy from four parameters alone (phase transition strength, wall velocity and the speed of sound in both phases). 

We present a simple model that allows to vary the speed of sound in the broken phase and in the symmetric phase at will (which we call template model, Section~\ref{ssec:numu}). We compare the exact numerical results in specific models with the results obtained in the template model that is obtained by matching the strength parameter and the two speeds of sound. We generally find very good agreement (Section~\ref{ssec:Compare}). We also study to what extend a sizable deviation in the speed of sound is possible (Section~\ref{sec:SMsinglet}). In the appendices, we present some more benchmark models and a Python code snippet that can be used to calculate the energy fraction in the  template model.

\section{Hydrodynamics}\label{sec:hydro}
In this section, we first lay out the hydrodynamic equations that need to be solved. We then identify which parts of the analysis are model dependent and show how this dependence can actually be captured in three parameters. Then we demonstrate how we can use a simplified equation of state to compute the kinetic energy fraction as a function of only these three parameters. We then compare with other methods in the literature. In Section \ref{ssec:profiles} we show how the hydrodynamic solutions and the efficiency factor depend on the speed of sound.

\subsection{Hydrodynamic equations}\label{ssec:hydro}
We give a brief summery of the hydrodynamic equations that one needs to solve to find the kinetic energy fraction. For more details on the hydrodynamics, we refer to Refs.~\cite{LandauLifshitz, Kamionkowski:1993fg, KurkiSuonio:1995pp, Espinosa:2010hh,Giese:2020rtr}.

We describe the plasma as a perfect fluid, characterized by the thermodynamic quantities (internal) energy density $e$, pressure $p$ and enthalpy density $w$. The pressure is given by $p = -\mathcal F$, with $\mathcal F$ the free energy density, or temperature-dependent effective potential. The expression for $\mathcal F$ depends on the model and is determined by analyzing the particle physics model at finite temperature. The quantities $e$ and $w$ can be obtained from $p$ by
\be
	e \equiv T \frac{\partial p }{\partial T} - p \, ,\qquad w \equiv T \frac{\partial p}{\partial T} = p+e\, .
\ee
A related and important quantity is the speed of sound, which is defined as
\be
	c_s^2 \equiv \frac{dp/dT}{de/dT}\, .
\ee
For a relativistic plasma, the sound speed is $c_s^2 = 1/3$, but for a general equation of state, the speed of sound is temperature dependent. We will denote the speed of sound in the symmetric (broken) phase by $c_{s,s}$ ($c_{s,b}$). 

The energy-momentum tensor of the plasma is given by
\be
	T^{\mu\nu} = u^\mu u^\nu w + g^{\mu\nu} \, p \, ,
\ee
where $u^\mu$ denotes the four-velocity of the fluid and $g^{\mu\nu}$ the inverse Minkowski metric.

In order to obtain the hydrodynamic equations, we project the continuity equations \mbox{$\partial_\mu T^{\mu\nu} =0$} onto the directions parallel and perpendicular to the fluid flow, and assume that the system is self-similar, \emph{i.e.} depending only on the coordinate $\xi=R/t$, where $R$ is the radius of the bubble and $t$ the time since nucleation. The hydrodynamic equations can then be written as 
\bea
\frac{dv}{d\xi} &=&  \frac{2v(1-v^2)}{\xi(1-v\xi)}\left(\frac{\mu(\xi,v)^2}{c_s^2} -1 \right)^{-1} \, \nonumber,\\ 
\frac{dw}{d\xi} &=& w \left(1+ \frac{1}{c_s^2} \right) \gamma^2 \mu(\xi ,v) \frac{dv}{d\xi} \, , \label{eq:hydro}
\eea
where $v(\xi)$ is the fluid velocity, $\gamma$ the Lorentz factor $1/\sqrt{1-v^2}$ and 
\begin{equation}
	\mu(\xi,v) = \frac{\xi-v}{1 - \xi v}\, ,
\end{equation}
the boosted velocity.

Finding the solution to the hydrodynamic equations requires a set of boundary conditions. The boundary condition for $v$ is obtained by integrating the continuity equations in the wall frame across the bubble wall
\bea
\frac{v_+}{v_-}&=&\frac{e_b(T_-)+p_s (T_+)}{e_s (T_+)+p_b (T_-)} \, , \nonumber \\
v_+ v_-&=& \frac{p_s(T_+)-p_b(T_-)}{e_s(T_+)-e_b(T_-)} \, .
\label{eq:matching}
\eea
The subscript $+$ ($-$) is used for quantities right in front of (behind) the bubble wall, in the symmetric (broken) phase. The subscript $s$ ($b$) is used for thermodynamic quantities in the symmetric (broken) phase. Note that the velocities $v_\pm$ are defined in the frame where the bubble wall is at rest. Eq.~(\ref{eq:matching}) looks like two equations with four variables, but for a given wall speed, either $v_+$ or $v_-$ is always known. The value of $T_+$ is chosen such that the plasma at rest away from the bubble sits at the nucleation temperature $T_n$ (this might require a shooting algorithm, see the comments at the end of Section \ref{ssec:newalpha}). The two matching equations are then used to find the remaining velocity and $T_-$. 

The hydrodynamic equations allow for three type of solutions: deflagrations, hybrids and detonations. For fast-moving bubbles, the plasma is at rest in front of the bubble wall and forms a rarefaction wave behind it; this solution is called a detonation. For wall velocities smaller than the speed of sound (in the broken phase) the solution is a deflagration. In this case, the plasma forms a shock in front of the bubble wall, and is at rest behind it. For a wall velocity larger than the speed of sound in the broken phase, but smaller than the so-called Jouguet velocity (see Appendix A of Ref.~\cite{Giese:2020rtr})
\begin{equation}
	\xi_J = \frac{1+ \sqrt{3 \alpha_{\bar \theta}(1 -c_s^2 + 3 c_s^2 \alpha_{\bar \theta})}}{1/c_s + 3 c_s \alpha_{\bar\theta}}\, ,
\end{equation}
 the solution is a hybrid, consisting of a shock front \emph{and} a rarefaction wave. In Section \ref{ssec:profiles} the hydrodynamic solutions for these three cases are shown.

After solving the hydrodynamic equations, the kinetic energy of the plasma, $\rho_\text{fl}$, is obtained by integrating over the fluid profile
\be
	\rho_\text{fl} = \frac{3}{\xi_w^3} \int d\xi \xi^2 v^2 \gamma^2 w\, ,
\ee
with $\xi_w$ being the wall velocity.\footnote{Note that also for expansion modes that display a shock in front of the bubble wall, the kinetic energy is normalized to the volume that already released energy into the plasma, determined by $\xi_w$.} Our quantity of interest is $K=\rho_\text{fl}/e_n$, often called the kinetic energy fraction. $e_n$ is the energy density in the symmetric phase at the nucleation temperature.\footnote{
Note that, in Ref.~\cite{Giese:2020rtr}, we were interested in $K = \rho_\text{fl}/e_+$. For the detonation solutions that we were considering $e_+ = e_n$ holds, but this does not hold in the more general case of the present work.
}

\subsection{Model dependence in the hydrodynamic equations}\label{ssec:newalpha}

A quick look at Eq.~(\ref{eq:hydro}) reveals that the hydrodynamic equations depend on the equation of state solely through the speed of sound $c_s^2$. Further model dependence enters through the matching conditions (\ref{eq:matching}) for the velocity, and the boundary condition of $w$. We will now describe how the model dependence can be captured by a small number of parameters.

Let us first focus on the matching conditions for the velocity, Eq.~(\ref{eq:matching}). In Ref.~\cite{Giese:2020rtr} we introduced a new phase transition strength parameter
\be
	\aNEW_+ \equiv \frac{D\bar \theta(T_+)}{3w_+}  \, , \quad 
	 {\rm with} \quad \bar \theta \equiv e - \frac{p}{c_{s,b}^2} \, , 
\, \label{eq:def:pt} 
\ee
where the quantity $DX(T_+)$ is defined as
\be
	DX(T_+) = X_s(T_+) - X_b(T_+),
\ee
with $X = e,p,w$. 

Assuming that the temperature does not vary strongly, the matching conditions at the bubble wall can be expressed in terms of this phase transition strength $\aNEW_+$ and the speed of sound in the broken phase only
\be
\frac{v_+}{v_-} \simeq \frac{(v_+v_-/c_{s,b}^2-1) + 3\aNEW_+}{(v_+v_-/c_{s,b}^2-1) + 3 v_+v_- \aNEW_+} \, ,
\label{eq:omatchnew}
\ee
which can be solved for $v_+$ or $v_-$. The model dependence of the velocity matching conditions is thus captured by $\aNEW_+$ and $c_{s,b}$. 

If we make the additional assumption that the sound speed depends only weakly on the temperature, it can be approximated by its value at the nucleation temperature, and thus the model dependence of the hydrodynamic equations is captured by just the two numbers $c_{s,b}$ and $c_{s,s}$ (note that the latter does not enter for detonations).

Finally, the model dependence in the matching of $w$ can be removed by working with $w(T)/w_+$ instead of $w(T)$. As a result, we do not determine the kinetic energy $\rho_\text{fl}$ (since it is model dependent), but instead we can determine the `efficiency factor'
\be
	\kappa = \frac{4\rho_\text{fl}}{D\bar \theta} = \frac{4\rho_\text{fl}}{3\aNEW_+ w_+}\, ,\label{eq:kappaK}
\ee
from which it is easy to obtain $K$.

Before showing how this projection onto three model-dependent quantities simplifies the hydrodynamic computations, two comments are in order. First, in Ref.~\cite{Giese:2020rtr}, where $\aNEW$ was introduced, we focused on detonations, for which the temperature in front of the bubble wall is equal to the nucleation temperature, \emph{i.e.} $T_+ = T_n$. $\aNEW_+$ could thus be computed from the particle physics model, without knowledge of the hydrodynamics. For deflagrations and hybrids, the temperature which is relevant for the matching, $T_+$, is not equal to the temperature $T_n$. It is thus necessary to find the relation between the two temperatures, or equivalently between $\aNEW_+$ and $\aNEW_n$, defined as
\be
	\aNEW_n = \frac{D \bar\theta(T_n)}{3w_n}\, \label{eq:anewn}.
\ee
It turns out that this relation is model independent, in the sense that $\aNEW_n$ and the sound speeds are still sufficient to fix the hydrodynamics. Still, solving the hydrodynamics for a given $\aNEW_n$ is more challenging for deflagrations and hybrids than it was for detonations, as a shooting method is required to find the value of $\aNEW_+$ (which enters the matching) that results in the required value of $\aNEW_n$. The relevant efficiency factor becomes
\be
	\kappa = \frac{4 \rho_\text{fl}}{3 \aNEW_n w_n}\,\label{eq:KappaKn} .
\ee 

Second, in the detonation case, the only relevant sound speed was the one of the broken phase. Although the sound speed in the symmetric phase enters in the hydrodynamic equations for deflagrations and hybrids, the phase transition strength $\aNEW_n$ and the velocity matching equation always depend on the speed of sound in the broken phase only. The reason is that Eq. (\ref{eq:omatchnew}) is obtained by expanding the thermodynamic quantities around the symmetric phase, see Ref.~\cite{Giese:2020rtr} for further details.

\subsection{Mapping onto the template model}\label{ssec:numu}

Now that we have shown that the model dependence can be captured by just $\aNEW$, $c_{s,s}$ and $c_{s,b}$, we need a method to compute the kinetic energy fraction $K$ that depends only on those three parameters and the wall velocity. This can be done by using a simple equation of state, introduced in Ref.~\cite{Leitao:2014pda}, that we will refer to as the  template model\footnote{In Ref.~\cite{Giese:2020rtr} we used the same model, but called it the $\nu$-model and we set $\mu=4$, which is sufficient for detonations.}
\begin{alignat}{3}
&p_s=\frac13 a_+ T^\mu -\epsilon \, ,& \qquad  & 
e_s= \frac13 a_+ (\mu-1) T^\mu + \epsilon\, , \nonumber \\ 
&p_b=\frac13 a_- T^\nu \, ,& \qquad &
e_b= \frac13 a_- (\nu-1) T^\nu \,,
\end{alignat}
where the parameters $a_+$ and $a_-$ are proportional to the number of relativistic degrees of freedom in the symmetric and broken phase respectively. Here, $\epsilon$ is the temperature-independent vacuum energy that is released in the phase transition. The parameters $\mu$ and $\nu$ are related to the sound speeds in the symmetric and broken phase, respectively:
\be
\label{eq:cs2}
	\mu=1+\frac{1}{c^2_{s,s}} \,,\qquad\, \nu=1+\frac{1}{c^2_{s,b}}. 
\ee
As discussed in Ref.~\cite{Giese:2020rtr}, after fixing $\aNEW_+$, $\mu$ and $\nu$, only one free parameter remains. A possible choice is
\be
	\chi = \frac{a_-}{a_+} T_{\rm cr}^{\nu-4}\, ,
\ee 
where $T_{\rm cr}$ is the critical temperature. We will consider different values of $\chi$ as corresponding to different `models'. As the matching equations (\ref{eq:omatchnew}) are exact for this equation of state and the sound speeds are constant, the value of $\kappa$ 
 is indeed exactly model independent, \emph{i.e.} does not depend on the value of $\chi$. This was explicitly demonstrated in Ref.~\cite{Giese:2020rtr}.
 
Our model-independent, simplified method to obtain the kinetic energy fraction in any model of new physics is as follows. Determine $\aNEW_n$, $c_{s,s}$ and $c_{s,b}$ for the particle physics model of interest, all evaluated at the nucleation temperature. Solve the hydrodynamic equations in the  template model to obtain $\kappa$. For the convenience of the reader, we provide a Python snippet in Appendix~\ref{App:code} that does this computation. As a last step, $K$ is determined through the relation~(\ref{eq:KappaKn}). The method is the same as the one proposed in Ref.~\cite{Giese:2020rtr}, but has now been extended to include hybrids and deflagrations.

\subsection[Other methods to determine $K$]{Other methods to determine \boldmath{$K$}}\label{ssec:othermethods}
In the literature, the main method to estimate $K$ without solving the hydrodynamic equations, is by mapping the model onto the bag equation of state
\begin{alignat}{3}
\label{eq:bagP}
&p_s = \frac13 a_+ T^4 - \epsilon \, ,& \qquad &
e_s = a_+ T^4 + \epsilon \, , \nonumber \\
&p_b = \frac13 a_- T^4 \, ,& \qquad &
e_b = a_- T^4 \, .
\end{alignat}
This equation of state can be obtained from the template model in the limit $c_{s,s}^2,c_{s,b}^2 \rightarrow  1/3$. The hydrodynamic analysis was done in Ref.~\cite{Espinosa:2010hh}, and a fit of $\kappa$ was given in terms of the wall velocity and $\alpha_\epsilon$, with $\alpha_\epsilon$ given by
\be
	\alpha_\epsilon = \frac{\epsilon}{a_+ T_n^4} = \frac{4 \epsilon}{3 w_n} \, .
\ee
For models that do not follow this equation of state, one needs to define a generalized phase transition strength to use the result of Ref.~\cite{Espinosa:2010hh} in an arbitrary model. The three most-used definitions are (we do not put a subscript $n$ on these $\alpha$'s, as they are always evaluated at the nucleation temperature)
\be
	\alpha_\theta = \frac{D\theta}{3w_n}\, ,\quad \alpha_p = -\frac{4Dp}{3w_n}\, , \quad \alpha_e = \frac{4De}{3w_n}\, ,
\ee
with
\be
	\theta = e - 3 p\, ,
\ee
the trace of the energy momentum tensor. Note that only $\alpha_\theta$ reduces to $\alpha_\epsilon$ for the bag model. It was demonstrated in Ref.~\cite{Giese:2020rtr} that these $\alpha$'s do not give a model-independent description of the template model, \emph{i.e.} after fixing $\alpha_\theta, \alpha_p$ or $\alpha_e$ some dependence on the value of $\chi$ remains.

\subsection{Comparison of different methods to compute the kinetic energy fraction}\label{ssec:Compare}

We will now compare our new method to compute the kinetic energy fraction, described in Sec.~\ref{ssec:numu}, to the other methods used in the literature. We will apply these methods to two benchmark models, the details of which are given in Appendix~\ref{appendix:benchmarks}. In total, we will use six different methods to compute $K$. They are summarized in Table~\ref{tab:methodComparison}, and explained below.

\begin{table}[h]
	\begin{center}
		\begin{tabular}{ | c  | c |}
			\hline
			M$_1$ & $K$ \\  
			\hline
			M$_2$ & $\left(\frac{D\bar\theta}{4e_n}\right) \kappa (\aNEW_n,c_{s,s},c_{s,b})\rvert_{\mu\nu}$ \\
			\hline
			M$_3$ & $\left(\frac{D\theta}{4e_n}\right) \kappa(\alpha_\theta)\rvert_{\rm bag}$ \\
			\hline
			M$_4$ & $\left(\frac{\alpha_\theta}{\alpha_\theta+1}\right) \kappa(\alpha_\theta)\rvert_{\rm bag}$ \\
			\hline
			M$_5$ & $\left(\frac{\alpha_p}{\alpha_p+1}\right) \kappa(\alpha_p)\rvert_{\rm bag}$ \\
			\hline
			M$_6$ & $\left(\frac{\alpha_e}{\alpha_e+1}\right) \kappa(\alpha_e)\rvert_{\rm bag}$ \\
			\hline
		\end{tabular}
		\caption{\small\label{tab:methods} Different methods to determine the kinetic energy fraction $K=\rho_{\rm fl}/e_n$ by mapping to the template model or bag model.}
		\label{tab:methodComparison}
	\end{center}
\end{table}

Method 1 is a fully numerical solution of Eq.~(\ref{eq:hydro}), without additional simplifications. The full temperature-dependent sound speed is used. Method 2 maps the model onto the template model, as described in Sec. \ref{ssec:numu}. $\kappa$ is determined as a function of $\aNEW_n, c_{s,s}$ and $c_{s,b}$ and the kinetic energy fraction is found via Eq.~(\ref{eq:KappaKn}). Methods 3--6 are often used in the literature. In Method 3 and 4 the model is mapped onto the bag equation of state, with $\alpha_\theta$ as the phase transition strength. In Method 4 the prefactor is furthermore simplified to $\alpha_\theta/(\alpha_\theta+1)$, an approximation which relies on the bag equation of state. In particular, this assumes that the adiabatic index is given by $\Gamma=w_n/e_n\simeq 4/(1+\alpha_\theta)/3$. Method 5 and 6 use the bag equation of state for $\alpha_p$ and $\alpha_e$ respectively.

\begin{figure}[h!]
	\centering
	\includegraphics[width=1\textwidth]{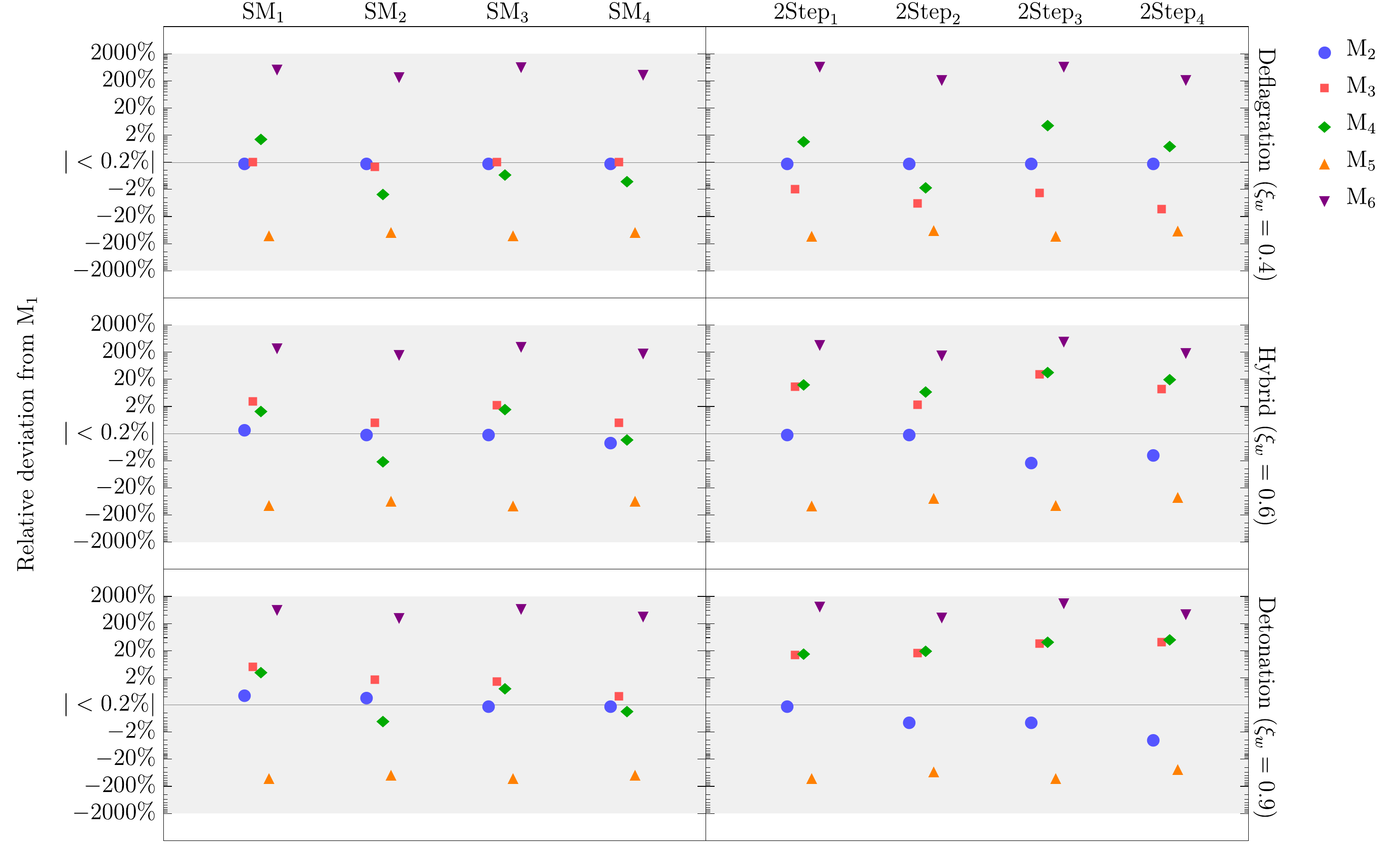}
	\caption{\small{Relative deviation from solving the exact hydrodynamic equations for the different approximation schemes M$_2$ to M$_6$. We show the deviation for several benchmark scenarios, four SM-like, and four 2Step phase transitions. The wall velocity is treated as an external parameter and is set to (0.4, 0.6, 0.9) in the (first, second, third) row. The exact definition of the benchmark scenarios can be found in Appendix \ref{appendix:benchmarks}.} }
	\label{fig:benchmarkmodels}
\end{figure}

We compute the kinetic energy fraction in all methods for a wall velocity $\xi_w$ of $0.4, 0.6$, and $0.9$, corresponding to a deflagration, hybrid and detonation respectively. The results are summarized in Figure \ref{fig:benchmarkmodels} and in Table~\ref{tab:final} in Appendix~\ref{appendix:benchmarks}, where the results of Methods 2--6 are given as a deviation from Method 1. In all cases Method 2 finds the smallest deviation from the full result. In most cases the deviation is $<1\%$, with the exception of the cases with relatively large $\aNEW_n$ ($\sim \mathcal O(0.1)$). The prevalence of Method 2 over the other methods is most apparent for hybrids and detonations. In Sec.~\ref{ssec:profiles} we study how the sound speed affects the different types of solutions. Method 3 and 4 perform reasonably, with deviations up to $\sim \mathcal O (50 \%)$. Method 5, based on the pressure difference, consistently underestimates $K$ by $50 - 100\%$, and method 6, based on the energy difference, overestimates the result by a few times $100\%$.

\begin{figure}[h!]
	\centering
	\includegraphics[width=0.9\textwidth]{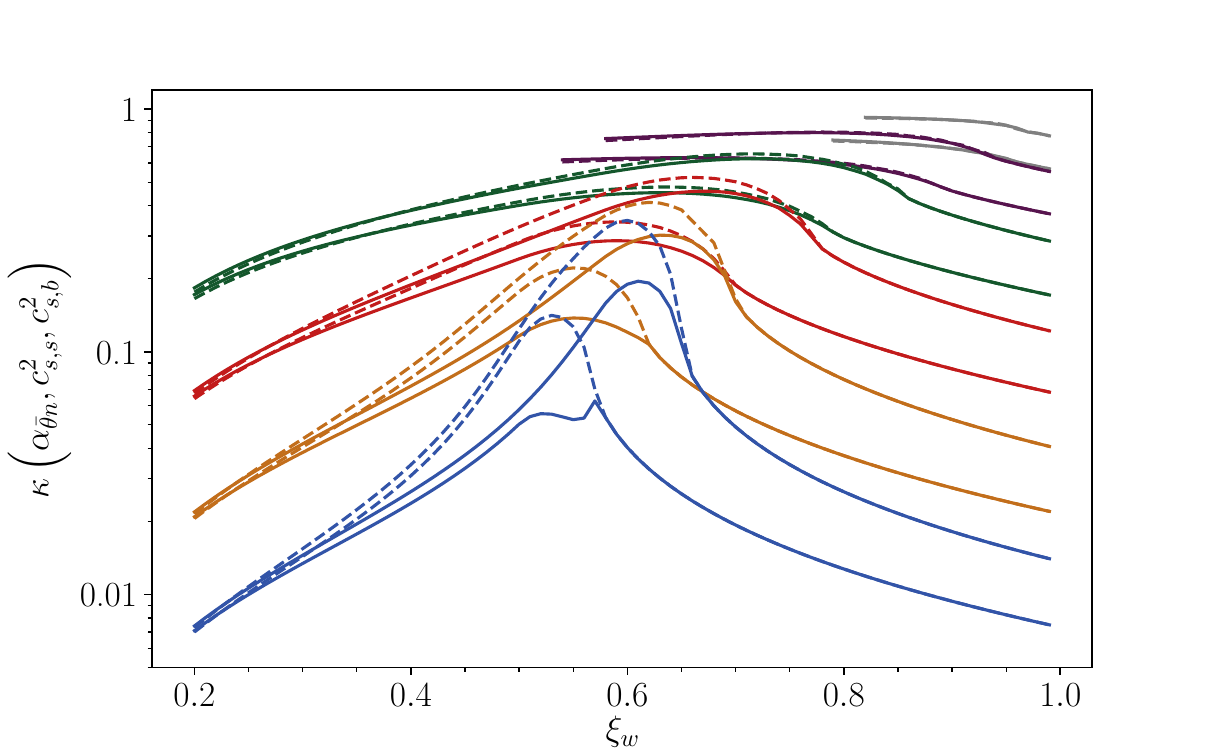}
	\caption{\small Efficiency factor for the template model. The different colors correspond to different values of $\aNEW_n={0.01,0.03,0.1,0.3,1,3}$. For each color, the upper line has $c^2_{s,b}=1/3$ and the lower line $c^2_{s,b} = 1/4$. The solid lines correspond to $c^2_{s,s} = 1/3$ and dashed lines correspond to $c^2_{s,s}=1/4$. }
	\label{fig:efficiencyfactor}
\end{figure}

\subsection{Sound speed corrections to velocity and enthalpy density profiles}\label{ssec:profiles}
In this section, we use the template model to understand how lowering the sound speed affects the different kind of solutions to the hydrodynamic equations. Figure~\ref{fig:efficiencyfactor} displays the effect of varying sound speeds on the efficiency factor for different values of $\aNEW_n$. It shows that the effect of varying the sound speeds is largest for detonations and hybrids. Detonations are unaffected by a change in $c_{s,s}$. In the paragraphs below we look at individual velocity and enthalpy density profiles for $\aNEW_n = 0.1$.

\begin{figure}[t!]
	\centering
	\includegraphics[width=1\textwidth]{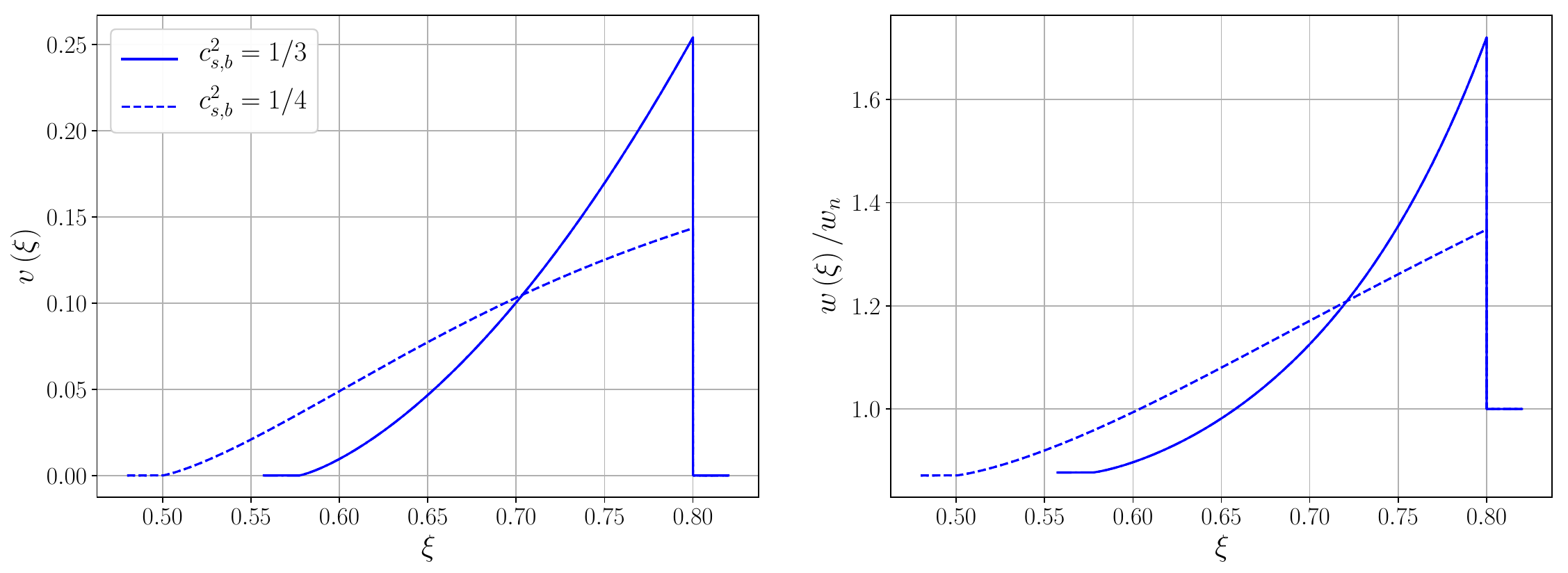}

	\centering
	\includegraphics[width=1\textwidth]{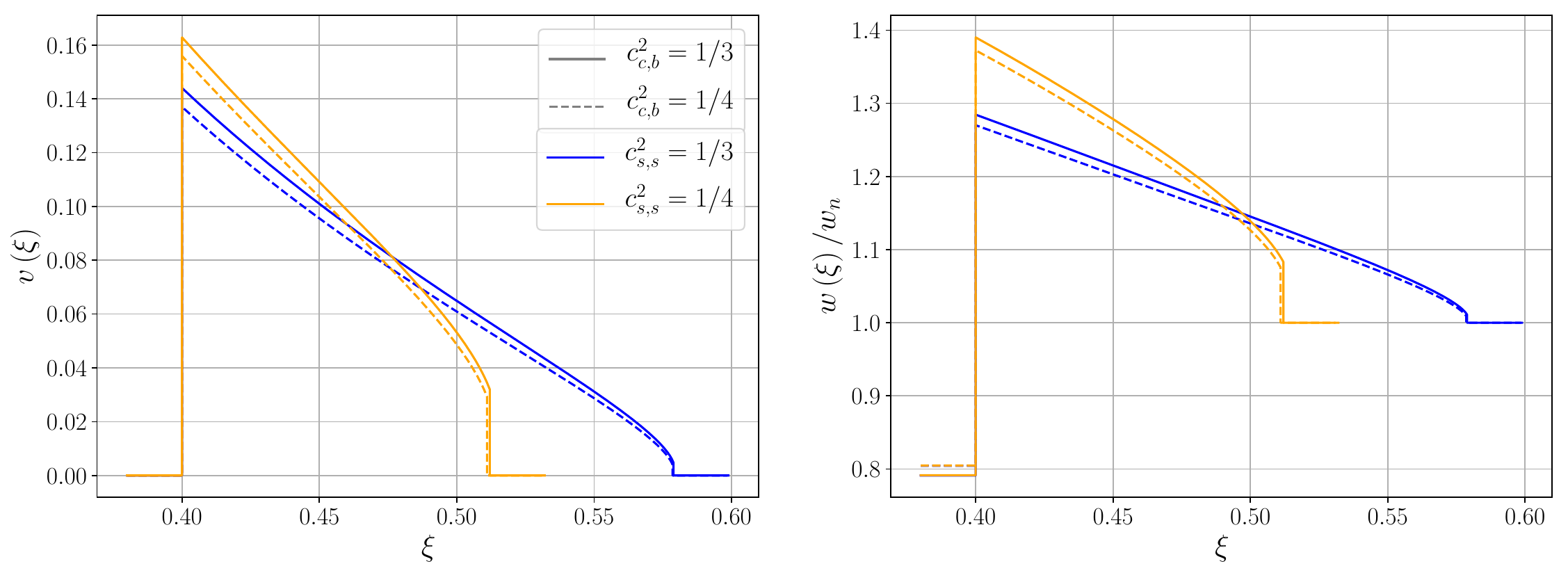}
	\centering
	\includegraphics[width=1\textwidth]{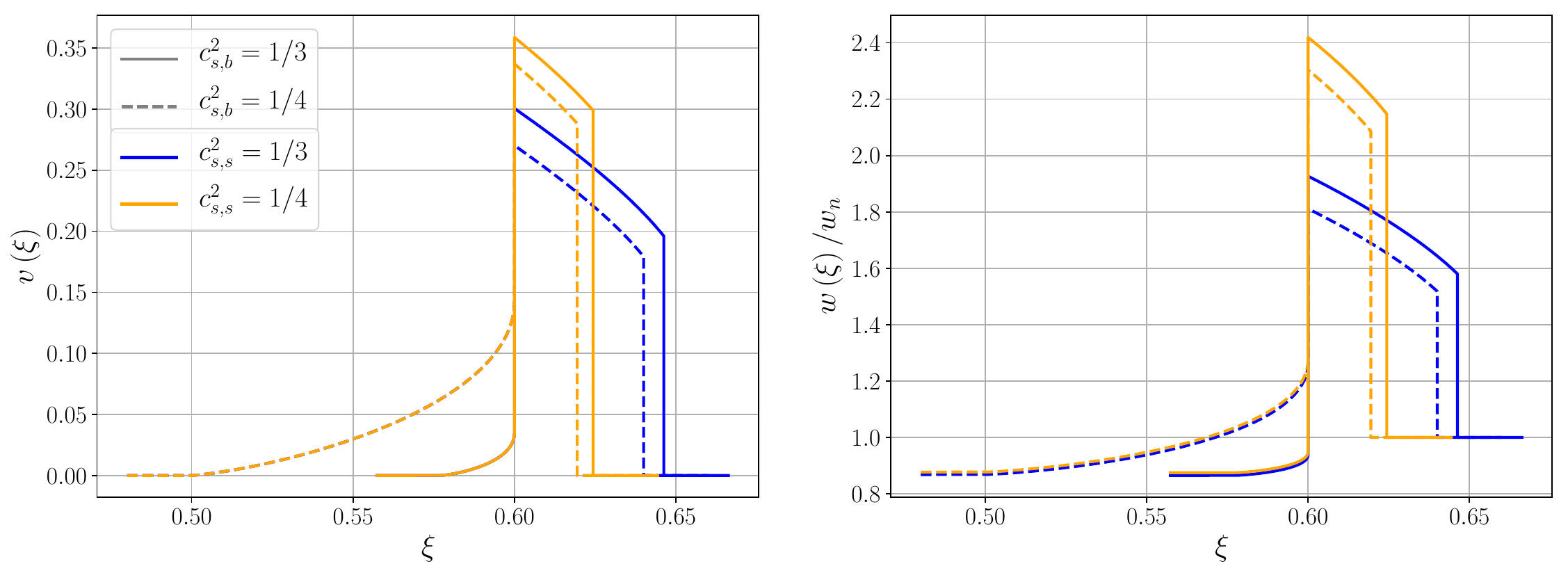}
	\caption{Velocity profiles (left column) and enthalpy density profiles (right column). The three rows correspond to detonations ($\xi_w=0.8$), deflagrations ($\xi_w=0.4$), and hybrids ($\xi_w=0.6$). The phase transition strength at nucleation temperature is set to $\aNEW_n=0.1$.}
\vskip 0.2cm
	\label{fig:detonationProfile}
	\label{fig:deflagrationProfile}
	\label{fig:hybridProfile}
\end{figure}

\paragraph{Detonations}
For detonation solutions the plasma in front of the wall is at rest, such that $v_+=\xi_w$ in the wall frame. Since the fluid velocity in the symmetric phase is zero, the sound speed in the symmetric phase never enters the hydrodynamic equations. The first row of Figure~\ref{fig:detonationProfile} showcases the velocity and enthalpy density profiles for detonations with wall velocity $\xi_w = 0.8$ and $\aNEW_n = 0.1$. Decreasing the speed of sound decreases the value of $v_-$ and consequently $w_-$ but horizontally stretches the profile (as it goes to zero at $\xi = c_{s,b}$). Overall, a decrease in sound speed in the broken phase leads to a suppression of the kinetic energy fraction, as can be seen in Figure~\ref{fig:efficiencyfactor}. 

\paragraph{Deflagrations}
For wall velocities smaller than the speed of sound in the broken phase $\xi_w < c_{s,b}$, the solution is a deflagration, with the plasma at rest behind the bubble wall ($v_- = \xi_w$). The profile now depends on both sound speeds, since $c_{s,s}$ enters in the hydrodynamic equations and the velocity matching depends on $c_{s,b}$. The second row of Figure~\ref{fig:deflagrationProfile} shows the velocity and enthalpy density profiles for a deflagration with $\xi_w = 0.4$. It becomes clear from the graph that decreasing the sound speed of the broken phase (slightly) decreases $v_+$. Figure~\ref{fig:efficiencyfactor} shows that this corresponds to a decrease in the efficiency factor. Decreasing the sound speed in the symmetric phase actually enhances the maximum value of the velocity and enthalpy density. This leads to an enhancement in the efficiency factor, as is clear from Figure~\ref{fig:efficiencyfactor}. This explains why Method 3 and 4 typically underestimate the kinetic energy fraction for deflagrations in the two-step model, where $c_{s,s}^2 < 1/3$ (see the first row in Figure \ref{fig:benchmarkmodels}).

\paragraph{Hybrids}
For wall velocities between the sound speed in the broken phase and the Jouguet velocity, the hydrodynamic solution is a hybrid, consisting of a shock front and a rarefaction wave. The velocity in the broken phase is given by $v_- = c_{s,b}$, which explains the fact that rarefaction waves in the third row of Figure~\ref{fig:hybridProfile} only depend on $c_{s,b}$. In the well-studied case of the bag model, hybrids are always supersonic deflagrations. For $c_{s,b} < c_{s,s}$ a hybrid can also be a subsonic detonation, in the sense that the wall velocity is smaller than the speed of sound in the symmetric phase. This possibility was pointed out in Ref.~\cite{Leitao:2014pda} and we also observe these modes here. The effect of lowering the sound speed in the symmetric phase is the same as for the deflagration case. Interestingly, for the rarefaction wave, the effect can be opposite to the detonation case. Decreasing the speed of sound in the broken phase can actually enhance the fluid velocity inside the bubble, and in turn suppress the velocity in the shock front. Looking at Figure~\ref{fig:efficiencyfactor}, the overall effect of lowering $c_{s,b}$ is to decrease the efficiency factor, whereas a decrease in $c_{s,s}$ enhances the efficiency factor. 

\section{Gravitational waves in the SM plus a singlet}\label{sec:SMsinglet}

We now apply our new method to determine the kinetic energy fraction to the full computation of the gravitational wave spectrum for a realistic model of new physics, the Standard Model (SM) extended by a real scalar singlet with $\mathbb Z_2$-symmetry. In this extension of the Standard Model, the phase transition can proceed in two steps. In the first step, the singlet obtains a vacuum expectation value and in the second step the electroweak symmetry gets broken. In this scenario, the phase transition can be rather strong, leading to a potentially observable gravitational wave signal. This setup has been studied in detail (see e.g.~Refs.~\cite{Espinosa:2011ax,Kurup:2017dzf, Kozaczuk:2015owa} and references therein).  

Note that the two-step phase transition is also one of the benchmark models in Appendix \ref{appendix:benchmarks}. In that case, however, we use the high-$T$ expansion, neglect the Coleman--Weinberg and daisy resummation terms, and we do not restrict one of the scalar fields to be the SM Higgs boson, and we therefore have more freedom in the parameter choices. In addition, we set the nucleation temperature by hand, instead of solving the bounce equation. In this section we do not make these approximations, and this leads to smaller deviations from $c_s^2 = 1/3$. We will briefly discuss the circumstances in which we do expect a significant deviation in the sound speed.
\subsection{Model}
We extend the Standard Model by a scalar field $s$, which is a singlet under the SM gauge groups and thus only couples to the Higgs. The tree-level scalar potential is given by:
\begin{equation}
	V_\text{tree}(H,s) = - \frac{ \mu_h^2}{2} h^2 + \frac{\lambda_h}{4} h^4 - \frac{\mu_s^2}{2} s^2 + \frac{\lambda_s}{4} s^4 + \frac{ \lambda_{hs}}{4} h^2 s^2 + \Delta V_h \,,
\end{equation}
where $h$ denotes the radial component of the Higgs field and $\Delta V_h$ is chosen such that the potential energy equals zero in the zero-temperature electroweak minimum. We analyze the potential at one-loop level, following the approach of Ref.~\cite{Espinosa:2011ax}. The Coleman--Weinberg contribution is given by
\begin{equation}
	V_\text{CW}(h,s) = \frac{1}{64 \pi^2} \sum_\alpha N_\alpha M_\alpha^4(h,s) \left[\log{\frac{M_\alpha^2(h,s)}{Q^2}} -C_\alpha \right]\, ,
\end{equation}
where the sum runs over the top quark, $W$ and $Z$ bosons and $h$ and $s$. $M_\alpha$ are the tree-level masses and $N_\alpha$ counts the number of degrees of freedom and is given by $\{-12, 6, 3, 1, 1\}$, respectively. The constant $C_\alpha$ equals $3/2$ for scalars and fermions and $5/6$ for gauge bosons. We take the renormalization scale $Q$ to be  the top quark mass at zero temperature. We have further removed the Goldstone boson contribution according to Ref.~\cite{Espinosa:2011ax}; including the Goldstone bosons does not change the qualitative picture, but shifts the nucleation temperature by $\mathcal{O}(\mathrm{<1\ GeV})$. For the fully consistent addition of Goldstone degrees of freedom to the effective potential at one-loop, see e.g. the Appendix of Ref.~\cite{Delaunay:2007wb}. We add a counterterm potential
\begin{equation}
	V_\text{CT}(H,s) = - \frac{\delta \mu_h^2}{2} h^2 + \frac{\delta\lambda_h}{4} h^4 - \frac{\delta\mu_s^2}{2} s^2 + \frac{\delta\lambda_s}{4} s^4 + \frac{\delta \lambda_{hs}}{4} h^2 s^2 + \delta\Delta V_h\,, 
\end{equation}
and we choose the counterterms such that the tree-level structure of the potential is maintained (for details, also see Ref.~\cite{Espinosa:2011ax}). The finite-temperature correction is given by
\begin{align}
	V_T(h,s) = &\frac{T^4}{2\pi^2} \sum_\alpha N_\alpha \int_0^\infty dx x^2 \log{\left[1 \pm e^{-\sqrt{x^2 + M_\alpha^2(h,s)/T^2}} \right]} \nn \\
				 &+ \frac{T}{12\pi} \sum_{\text{bosons}\, \alpha} N_\alpha \left[M_\alpha^3(h,s) - M_{T,\alpha}^3(h,s,T) \right]\, ,\label{eq:VT}
\end{align}
where the $M_{T,\alpha}$ are the finite-temperature masses, computed in the high-temperature expansion. The first line is the standard one-loop thermal integral, where the positive sign is taken for fermions and the negative sign for bosons. The second line corresponds to the daisy resummation. It is absent for fermions and transverse gauge bosons. In order to correctly account for the field-independent contribution from all relativistic particle species, we add the contribution from all relativistic particles that we have not accounted for in Eq.~(\ref{eq:VT}) by hand:
\begin{equation}
	\delta V_T(h,s) = - \frac{\pi^2}{90} g_*' T^4 \,,\quad g_*' = \frac{345}{4} \, .
\end{equation}
The full effective potential is then given by
\be
\label{eq:Veff}
V_{\rm eff}(h,s,T)=V_\text{tree}+V_\text{CW}+V_\text{CT}+V_T+\delta V_T\, .
\ee


\subsection{Gravitational-wave signal}


Based on the effective potential in Eq.~\eqref{eq:Veff}, we are able to compute the expected GW signal for parameter points that result in a first-order phase transition.
In doing so, we will follow Ref.~\cite{Caprini:2019egz} and restrict ourselves to the contribution to the signal from sound waves.
The GW signal from bubble collisions is subdominant~\cite{Bodeker:2017cim,Ellis:2019oqb,Hoeche:2020rsg,Ellis:2020nnr,Vanvlasselaer:2020niz}, while the GW signal from magnetohydrodynamic turbulence currently still requires a better theoretical description.
Omitting the contribution from turbulence therefore corresponds to a conservative approach that results in a lower estimate of the actual GW signal.
The sound-wave contribution to the GW signal can be written as follows,
\begin{equation}
\label{eq:Omegasw}
\Omega_{\rm sw}\left(f\right) = \Omega_{\rm tot}\, \mathcal{S}\left(f/f_{\rm peak}\right) \,,
\end{equation}
where $\Omega_{\rm tot}$ denotes the total GW energy density sourced by sound waves, in units of the critical energy density $\rho_{\rm crit} = 3\,H_0^2\,M_{\rm Pl}^2$, and $\mathcal{S}$ characterizes the spectral shape of the signal.
A more detailed discussion of $\Omega_{\rm tot}$ and $\mathcal{S}$ can be found in Ref.~\cite{Schmitz:2020rag}; here, we will just summarize the most important points relevant for the present analysis.


The spectral shape is often approximated by a broken power law.
Ref.~\cite{Caprini:2019egz}, \textit{e.g}, uses an expression that is motivated by the numerical simulations in Refs.~\cite{Hindmarsh:2015qta,Hindmarsh:2017gnf}
\begin{equation}
\label{eq:S}
\mathcal{S}\left(x\right) = \frac{x^p}{\mathcal{N}\left(p,q,n\right)\left[q/\left(p+q\right) + p/\left(p+q\right) x^n\right]^{\left(p+q\right)/n}} \,,\quad \left(p,q,n\right) = \left(3,4,2\right) \,,
\end{equation}
where $\mathcal{N}$ ensures that the frequency integral over $\mathcal{S}$, on a logarithmic frequency scale, is normalized to one, $1/{\mathcal{N}\left(3,4,2\right)} \simeq 0.687$.
A broken power law typically provides a good description of the GW spectrum, unless the wall velocity is close to the Jouguet velocity.
In this case, a second scale emerges in the spectrum, such that $\mathcal{S}$ should be replaced by double broken power law~\cite{Hindmarsh:2016lnk,Hindmarsh:2019phv,Jinno:2020eqg}.
The frequency $f_{\rm peak}$ in Eq.~\eqref{eq:Omegasw} describes the position of the peak in the GW spectrum~\cite{Hindmarsh:2017gnf},
\begin{equation}
f_{\rm peak} \simeq 26\times10^{-3}\,\textrm{mHz}\:\bigg(\frac{z_{\rm peak}}{10}\bigg)\left(\frac{1}{R_*H_*}\right)\left(\frac{100}{g_s^*}\right)^{1/3}\left(\frac{g_\rho^*}{100}\right)^{1/2}\left(\frac{T_*}{100\,\textrm{GeV}}\right) \,.
\end{equation}
Here, $T_*$ denotes the temperature at the time of bubble percolation, which we will estimate by the temperature at the time of bubble nucleation $T_n$ in the following; $g_\rho^*$ and $g_s^*$ are the effective numbers of degrees of freedom contributing to the radiation energy and entropy densities at $T = T_*$, respectively; and $R_*H_*$ is the mean bubble separation, in units of the Hubble radius $H_*^{-1}$, at $T = T_*$.
The numerical coefficient $z_{\rm peak}$ controls the hierarchy between $R_*^{-1}$ and $f_{\rm peak}$ at the time of the phase transition and needs to be determined numerically.
A characteristic value is $z_{\rm peak} \simeq 10$~\cite{Hindmarsh:2017gnf}.


The bubble separation scale $R_*$ can be related to the duration of the phase transition, or nucleation rate parameter $\beta$, and the characteristic velocity scale of the expanding bubbles.
In the following, we will assume that this characteristic velocity scale corresponds to the velocity of the sound shell, $\xi_{\rm sh}$, which allows us to write
\begin{equation}
\label{eq:Rstar}
R_* \simeq \left(8\pi\right)^{1/3} \frac{\xi_{\rm sh}}{\beta} \,.
\end{equation}
In the case of detonations, $\xi_{\rm sh}$ coincides with the wall velocity, $\xi_{\rm sh} = \xi_w$, whereas for deflagrations, $\xi_{\rm sh}$ is given by the speed of sound in the symmetric phase, $\xi_{\rm sh} \simeq c_{s,s}$. 
These relations motivate us to estimate $R_*$ by the following rough expression,
\begin{equation}
R_* \simeq \left(8\pi\right)^{1/3} \frac{\max\left\{\xi_w,c_{s,s}\right\}}{\beta} \,.
\end{equation}
A more careful evaluation of $R_*$ would require a more precise determination of $\xi_{\rm sh}$ based on the hydrodynamic equations.
A more careful treatment is, in particular, also necessary in the case of hybrid transitions, where $\xi_{\rm sh} > \xi_w$.
The bubble nucleation parameter $\beta$ in Eq.~\eqref{eq:Rstar} is defined in terms of the temperature derivative of the three-dimensional Euclidean bounce action $S_3$,
\begin{equation}
\beta = H_* T_* \frac{d}{dT} \left.\frac{S_3\left(T\right)}{T}\right|_{T_*}  \, .
\end{equation}
We use the Mathematica package \textsc{FindBounce}~\cite{Guada:2018jek, Guada:2020xnz} to compute $S_3$ as a function of $T$ based on the effective potential in Eq.~\eqref{eq:Veff}.
The nucleation temperature $T_n$ is then determined by the condition that the probability of nucleating one bubble per Hubble volume and time approaches $P \simeq 1$. For phase transitions close to the electroweak scale this amounts to the requirement that (for a discussion of this criterion see e.g. Ref.~\cite{Caprini:2019egz})
\begin{equation}
\frac{S_3\left(T_n\right)}{T_n} \simeq 140 \,.
\end{equation}


\begin{table}
	\begin{center}
		\begin{tabular}{ | c  | c | c || c | c | c | c | c | c | c|}
			\hline
			$m_s (\text{GeV})$ & $\lambda_s$ & $\lambda_{hs}$ & $T_n (\text{GeV})$ & $\beta/H_*$ & $\alpha_e$ & $\aNEW_n$ & $c_{s,b}^2 $ & $c_{s,s}^2$ \\
			\hline\hline
			$300 $ & 1.90 & 3.50 & 87.3 & 288 & 0.070 & 0.035 & 0.324 & 0.333 \\
			$250 $ & 2.80 & 2.80 & 71.1 & 152 & 0.126 & 0.075 & 0.325 & 0.334 \\
			$250 $ & 0.40 & 2.26 & 98.9 & 367 & 0.051 & 0.022 & 0.325 & 0.333 \\
			$170 $ & 2.80 & 1.80 & 69.5 & 335 & 0.119 & 0.065 & 0.324 & 0.334 \\
			\hline
		\end{tabular}
		\caption{\small\label{tab:singlet} Phase transition parameters for four characteristic benchmark points in the $\mathbb{Z}_2$-symmetric real-scalar-singlet extension of the Standard Model.
		$m_s$ denotes the zero-temperature mass of the singlet in the electroweak vacuum.
		The numerical precision of the two sound speeds $c_{s,b}^2$ and $c_{s,s}^2$ is of $\mathcal{O}\left(10^{-4}\right)$.
		At this level of precision, $c_{s,s}^2$ is consistent with $c_{s,s}^2 = 1/3$, while $c_{s,b}^2$ deviates from the relativistic value by about two to three percent.}
		\label{tab:Z2singletParameters}
	\end{center}
\end{table}


The total GW energy density $\Omega_{\rm tot}$ in Eq.~\eqref{eq:Omegasw}, finally, is given by
\begin{equation}
\label{eq:Otot}
\Omega_{\rm tot} = \min\left\{1,H_*\tau_{\rm sh}\right\} \times 3\,F\,\tilde{\Omega}\,R_*H_*\,K^2 \,.
\end{equation}
Here, $F$ accounts for the redshift of the signal from $T = T_*$ to the present time,

\begin{equation}
F = \left(\frac{g_\rho^*}{g_\rho^0}\right)\left(\frac{g_s^0}{g_s^*}\right)^{4/3} \Omega_\gamma^0 \simeq 1.6 \times 10^{-5} \left(\frac{1}{h^2}\right)\left(\frac{g_\rho^*}{100}\right)\left(\frac{100}{g_s^*}\right)^{4/3}\,,
\end{equation}
where $h$, the dimensionless Hubble parameter in the present epoch, is defined via the relation $H_0 = 100\,h\,\textrm{km}/\textrm{s}/\textrm{Mpc}$.
$\tilde{\Omega}$ measures the efficiency of GW production from sound waves and follows from integrating the shear
stress unequal-time correlator of the bulk fluid.
As shown in Ref.~\cite{Hindmarsh:2017gnf}, this quantity is approximately constant in the case of weak phase transitions, $\tilde{\Omega} \sim 0.01$.
The prefactor in Eq.~\eqref{eq:Otot}, $\min\left\{1,H_*\tau_{\rm sh}\right\}$, accounts for the onset of shock formation in the plasma after some time $\tau_{\rm sh}$, which results in a suppression of the sound-wave signal~\cite{Ellis:2018mja,Ellis:2019oqb,Ellis:2020awk}, if shocks should appear within less than a Hubble time, $H_*\tau_{\rm sh} < 1$.~\footnote{
Ref.~\cite{Guo:2020grp} gives the following functional form for the suppression factor in the bag model
\begin{equation*}
	\Upsilon = 1 - \frac{1}{\sqrt{1 + 2H_* \tau_{\rm sh}}}\, .
\end{equation*}
Obtaining a better understanding of the suppression factor requires lattice simulations into the turbulent regime. 
}
The lifetime of the sound-wave source can be estimated in terms of the mean bubble separation $R_*$ and the enthalpy-weighted root-mean-square of the fluid velocity $\bar{U}_f$,
\begin{equation}
\tau_{\rm sh} \simeq \frac{R_*}{\bar{U}_f} \,,\quad \bar{U}_f = \left(\frac{K}{\Gamma}\right)^{1/2} \,,
\end{equation}
where $\Gamma= w_n/e_n$ is the mean adiabatic index of the plasma.
In the following, we will focus on parameter points where indeed $H_*\tau_{\rm sh} < 1$, such that the total GW energy density can be written as
\begin{equation}
\label{eq:OtotK}
\Omega_{\rm tot} = 3\,\Gamma^{1/2}F\,\tilde{\Omega}\left(R_*H_*\right)^2 K^{3/2} \,.
\end{equation}


 \begin{figure}[t]
	\centering
	\includegraphics[width=1\textwidth]{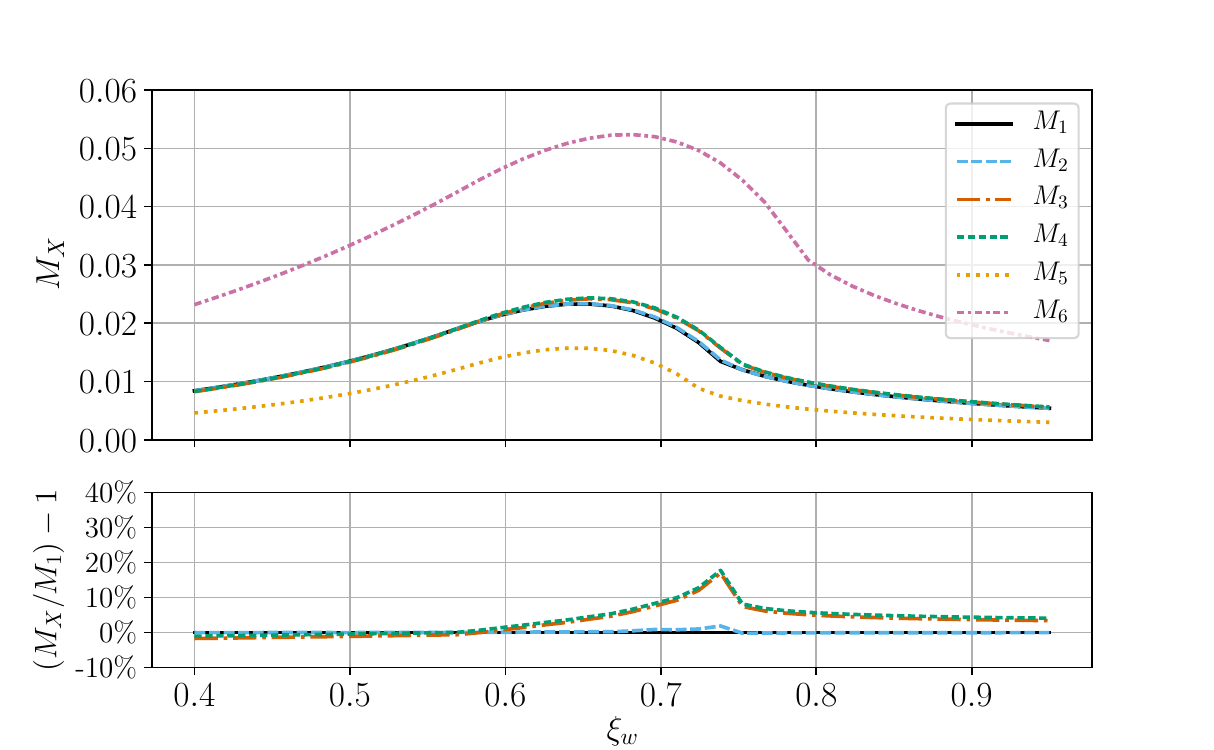}
	\caption{Upper panel: efficiency factor as a function of the wall velocity for the benchmark model with $m_s = 170$ GeV computed in all six methods. Lower panel: deviation in percent from Method 1. We only show Method 2--4; the deviations of Method 5 and 6 are much larger.}
	\label{fig:devEff}
\end{figure}


Eq.~\eqref{eq:OtotK} captures the relation between our hydrodynamic analysis in this paper and the implications for the GW signal: 
a change in the kinetic energy fraction by a factor $a$ results in a change of the GW amplitude $\Omega_{\rm tot}$, and hence of the expected signal-to-noise ratio (see Ref.~\cite{Schmitz:2020rag} for details), by a factor $a^{3/2}$.
In the following, it will therefore suffice to discuss the predictions for $K$ according to the six different methods listed in Table~\ref{tab:methodComparison}.
To do so, we consider the four characteristic benchmark points in Table~\ref{tab:singlet} to estimate the possible deviations from $c_{s,b}^2 = c_{s,s}^2 = 1/3$ that we may expect in the real-scalar-singlet extension of the Standard Model.
All four benchmark points in Table~\ref{tab:singlet} correspond to two-step phase transitions, where first the $\mathbb{Z}_2$ symmetry along the singlet direction becomes broken and then, in a second step, this symmetry becomes restored as the Higgs obtains a nonvanishing expectation value.
We find that, given the precision of our numerical analysis, $c_{s,s}^2$ is consistent with $c_{s,s}^2 = 1/3$, while the speed of sound in the broken phase typically deviates from the relativistic value by two to three percent.
For each benchmark point, we also compute $\aNEW_n$ and $\alpha_e$, which was proposed as the generalization of $\alpha_\epsilon$ in Ref.~\cite{Caprini:2015zlo}, and is used in many studies in the literature. As expected, $\aNEW_n$ and $\alpha_e$ deviate from each other quite significantly. 


As was already clear from Figure~\ref{fig:efficiencyfactor}, the difference between the efficiency factors computed in our new approach or using the bag equation of state depends sensitively on the wall velocity.
In Figure~\ref{fig:efficiencyfactor}, the sound speed was treated as a free parameter and could thus deviate quite strongly.
In Figure~\ref{fig:devEff}, we show the difference in the efficiency factors for the benchmark point with $m_s = 170$ GeV.
For the other benchmark points, we find comparable differences.
Around the Jouguet velocity, the difference in the value of $K$ between our method and the old methods is $\mathcal O(20\,\%)$, leading to a deviation of $\mathcal O(30\,\%)$ in the GW signal (but notice that also the lattice results of Ref.~\cite{Hindmarsh:2017gnf} somewhat deviate from the scaling Eq.~(\ref{eq:OtotK}) in this regime).
In the model under consideration, we never get strong deviations from $c_s^2 = 1/3$, which leads to only moderate differences between the new and old estimates of the GW signal.
In the next section, we point out sources for stronger deviations in the speed of sound.


\subsection{Deviations in the speed of sound}


In this section we try to give some insight on how deviations from $c_s^2 = 1/3$ occur. The sound speeds in the symmetric and broken phase at some temperature $T_0$ are given by
\begin{align}
c_{s,s}^2\left(T_0\right)&=\frac{1}{T}\frac{V_{\mathrm{eff}}'}{ V_{\mathrm{eff}}''}\bigg|_{\left(h=0,s=v_s(T_0),T=T_0\right)} \nonumber \, ,\\
c_{s,b}^2\left(T_0\right)&=\frac{1}{T}\frac{V_{\mathrm{eff}}'}{ V_{\mathrm{eff}}''}\bigg|_{\left(h=v_h(T_0),s=0,T=T_0\right)}.
\end{align}
Here, a prime denotes the total derivative with respect to $T$ and the scalar field values are evaluated in the minima $v_h$ and $v_s$ of the symmetric and broken phase respectively (the $\mathbb Z_2$-symmetry ensures that the EW symmetric minimum lies on the $h=0$ axis, and the EW broken minimum on the $s=0$ axis). The  minima are functions of temperature, $v_h(T)$ and $v_s(T)$,  and for the SM particle content their $T$ dependence is negligible\footnote{In the SM the crossover nature of the EW phase transition in fact should lead to a dip in the sound speed from $v_h'(T)\neq0$ near the crossover temperature as shown in Figure~7 of Ref.~\cite{DOnofrio:2015gop}. This effect is however expected to be smaller in first-order phase transitions and relevant near the critical temperature. It cannot compete with the reduction of the sound speed  induced by the explicitly $T$-dependent terms from Eq.~(\ref{eq:highTexp}). More importantly: in our analysis the input quantities are the sound speeds evaluated at the nucleation temperature. In case $T_n$ is too close to the critical temperature, the phase transition strength becomes too small to be seen in experiment. Thus the interesting scenarios have some degree of supercooling which usually is strong enough for $v_h(T_n)\simeq v_h(0)$ and $v_s(T_n)\simeq v_s(0)$ for setups with a particle content close to the SM.}, 
so the behavior of the sound speed depends predominantly on $V_T + \delta V_T$. We can understand the temperature-dependence of the sound speed better by expanding the thermal integral in Eq.~(\ref{eq:VT}) in the high-temperature limit $M_\alpha^2/T^2 \ll 1$.
\begin{align}
	&\frac{T^4}{2\pi^2} \sum_\alpha N_\alpha \int_0^\infty dx x^2 \log{\left[1 \pm e^{-\sqrt{x^2 + M_\alpha^2(h,s)/T^2}} \right]} \, \nonumber \\
	 \sim & \sum_{{\rm bosons}\, \alpha} N_\alpha \left( -\frac{\pi^2 T^4}{90} + \frac{M_\alpha^2 T^2}{24} - \frac{M_\alpha^3 T}{12 \pi} \right) + \sum_{{\rm fermions}\, \alpha} N_\alpha \left( \frac{7 \pi^2 T^4}{720} - \frac{M_\alpha^2 T^2}{48} \right)\, .\label{eq:highTexp}
\end{align}
Massive particles thus contribute to a deviation in the sound speed from $c_s^2 = 1/3$ (note that the above expansion is not valid for particles with masses $M_\alpha^2 \gtrsim T^2$, which get Boltzmann suppressed). In our model under consideration, there is only a handful of massive particles, and we thus do not expect large deviations.

 \begin{figure}[t!]
	\centering
	\includegraphics[width=0.8\textwidth]{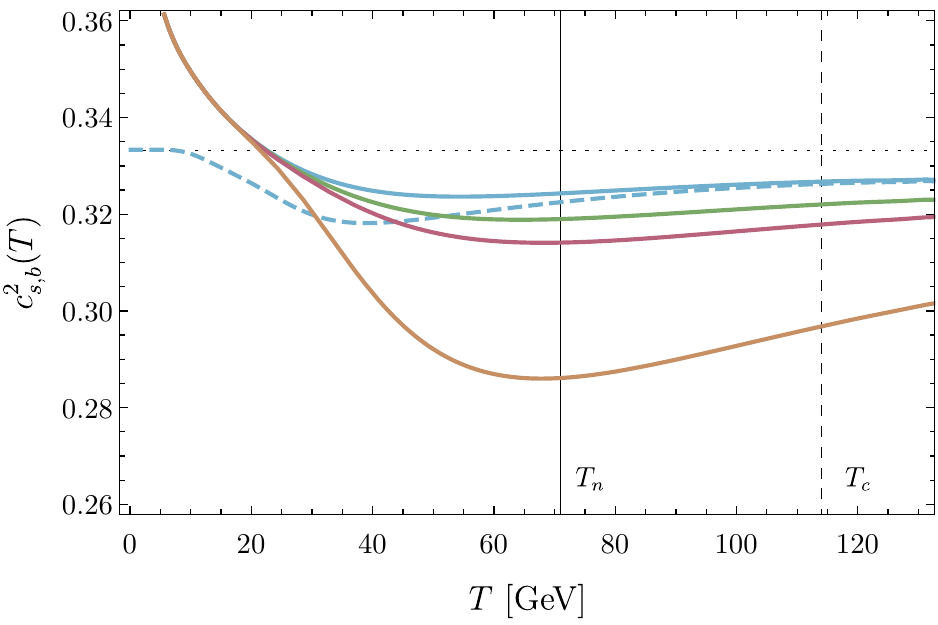}
	\caption{\small Sound speed in the broken phase as a function of the temperature. The blue curve assumes no extra particles, and the green, pink and orange lines assume $g_\pi=10, 20$ and $100$ additional scalar particles, respectively. The nucleation temperature is denoted by $T_n$ and is independent from $g_\pi$. $T_c$ denotes the critical temperature based on the criterion that the EW broken minimum must be favorable at nucleation temperature. For temperatures $T\rightarrow 0$ there is an IR divergence from the high-temperature expansion of the daisy terms. To quickly see that this is indeed the origin of the divergence one can  enforce Boltzmann-suppression by hand via multiplying the second line of Eq.~(\ref{eq:VT})  with a factor $\exp(-m_W^2/T^2)$, which reproduces the correct IR behaviour $c_s^2\rightarrow1/3$ (see dashed line). }
	\label{fig:cswithdarkmatter}
\end{figure}

Moreover, the daisy resummation in the second line of Eq.~(\ref{eq:VT}) suppresses deviations from $c_s^2 = 1/3$. The daisy resummation effectively replaces the term $M_\alpha^3 T/(12\pi)$ by
 $M_{T,\alpha}^3 T/(12\pi)$. Assuming that the $T$-dependent part dominates in $M_{T,\alpha}$, the contribution scales as $T^4$ and thus pushes the sound speed towards radiation and $c_{s}^2 = 1/3$. A more significant modification in the sound speed is thus expected in a model with additional massive fermions, which are not included in the daisy resummation, or weakly interacting bosons, for which $M_{T,\alpha} \sim M_\alpha$\,~. We realize the latter by adding a new type of extra non SM-charged scalar $\pi$ with number of degrees of freedom $g_\pi$, where $\pi$ is super weakly interacting (vanishing couplings to the SM), yet thermalized. The mass $m_\pi$ is then a free parameter and impacts the balance of non-relativistic to relativistic degrees of freedom in Eq.~(\ref{eq:highTexp}) while leaving the determination of $T_n$ untouched due to its non-interacting nature. This construction is the simplest way to induce $c_{s,s}^2<1/3$ and $c_{s,b}^2<1/3$ where the size of the deviation is controlled by $g_\pi$.

In Figure~\ref{fig:cswithdarkmatter} we show how adding extra scalars with $m_\pi=250\,{\rm GeV}$ affects the speed of sound in the broken phase. The sound speed gets lowered over a large range of temperatures, and one does not need to tune the nucleation temperature to observe this effect. Adding $g_\pi=20$ degrees of freedom corresponds to an increased relative deviation of $35\%$, at $\xi_w=0.74$, for M$_3$ (the most accurate approach found in the literature) relative to M$_1$; compared to the $15\%$ shown for $g_\pi=0$ in Figure~\ref{fig:devEff}. In both cases, the mapping to the template model is precise at percent level. The somewhat extreme addition of $g_\pi=100$ particles showcases that the assumption $c_{s}^2\sim const$ required by the template model can break down and larger deviations between our mapping M$_2$ and the exact calculation M$_1$ is the outcome, up to -12\%; compared to +75\% for M3. 

\section{Summary}\label{sec:summary}
%
We have presented a model-independent method to determine the fraction of energy going into kinetic energy of the fluid, $K$, in a cosmological first order phase transition. $K$ is an important quantity for the determination of the gravitational wave signal (see Eq.~(\ref{eq:OtotK})). We have shown that $K$ can be obtained from the efficiency factor $\kappa$ (see Eq.~(\ref{eq:KappaKn})), which is determined by solving the hydrodynamic equations. It turns out that the efficiency factor only depends on four parameters, the phase transition strength parameter $\aNEW_n$ (see Eq.~(\ref{eq:anewn})), the speed of sound in both phases and the bubble wall velocity (which we take as an external parameter), up to small corrections due to temperature dependence of the speed of sound. This work is a generalization of Ref.~\cite{Giese:2020rtr}, where the analysis was limited to detonations.

The above observations allow for an easy way to obtain the kinetic energy fraction in a specific model: one can measure the above parameters and match onto the template model. The hydrodynamic equations of the template model can then be solved using the code snippet of Appendix~\ref{App:code}. 

In Section~\ref{ssec:Compare} we compared our model-independent approach to the fully numerical computation of $K$. We also compared with other methods used in the literature, based on a mapping to the bag equation of state. The deviations between the full result and our new approach are typically sub-percent, with the exception of hybrids and detonations with a moderately strong phase transition ($\aNEW_n \sim \mathcal O (0.1)$), where the deviations are a few percent. In all cases, our method outperforms the other methods used in the literature. Mapping onto the bag equation of state using the phase transition strength from the pressure or energy density difference works particularly poorly, with differences of order $-(40 - 90 \%)$ (using $\alpha_p$) and $200 - 1000\%$ (using $\alpha_e$). Mapping onto the bag equation of state via the difference in the trace of the energy momentum tensor leads to smaller deviations.

In Section~\ref{sec:SMsinglet} we studied how our approach affects the gravitational wave signal in a two-step phase transition in the Standard Model extended by a singlet. Although the sound speed does not deviate strongly from $c_s^2 = 1/3$, we found that using our new approach still leads to a deviation in $K$ of up to $\mathcal O(20\%)$ compared to the old method with $\alpha_\theta$, corresponding to a $\mathcal O(30\%)$ difference in the gravitational wave signal. Using $\alpha_p$ or $\alpha_e$ leads to even stronger under- and overestimates of the gravitational wave signal. 

We have shown that the presence of weakly interacting massive degrees of freedom might affect the sound speed, leading to larger deviations in the gravitational wave signal. Identifying further models with a strong deviation in the sound speed will be the topic of future work.

\section*{Acknowledgements}
This project has been supported by the Deutsche Forschungsgemeinschaft under Germany's Excellence Strategy - EXC 2121 ``Quantum Universe'' - 390833306 (F.\,G., T.\,K. and J.\,v.d.V) and the European Union's Horizon 2020 Research and Innovation Programme under grant agreement number 796961, "AxiBAU" (K.\,S.).

\newpage

\appendix
\section{Benchmark models}
\label{appendix:benchmarks}

In this appendix we introduce the two toy models that are used for the comparison of the different methods to compute $K$ in Section~\ref{ssec:Compare}.

The first model is SM-like with a cubic term in the free energy density coming from thermal effects. The thermodynamic potentials in the symmetric and broken phases are given via the free energy density
\bea
{\cal F}(\phi,T) &=& -\frac{a_+}{3} T^4 + \lambda ( \phi^4 - 2 E \phi^3 T +\phi^2 (E^2 T_{\rm cr}^2 + d (T^2 - T_{\rm cr}^2)) ) \nn \\
&& \quad + \frac{\lambda}{4} (d-E^2)^2 T_{\rm cr}^4    \, ,
\eea
such that $p_s = -{\cal F}(0,T)$ and $p_b = -{\cal F}(\phi_{\rm min},T)$ with 
\be
\phi_{\rm min} = \frac34 E \, T + 
\sqrt{T^2 \, (9E^2/8 - d)/2 - \Tcr^2 \, (E^2 - d)/2} \, .
\ee
The model has four relevant parameters: $3\lambda/a_+$, $E$, $T_n/T_{\rm cr}$ and $d$, which in turn determine $\aNEW_n$ and $c_{s,b}$. The last term in the free energy density removes the cosmological constant at zero temperature in the broken phase. Symmetry breaking at low temperatures requires $d>E^2$. The barrier persists down to a temperature 
$T^2>T_{\rm cr}^2 (d-E^2)/(d-9E^2/8)$.

Table~\ref{tab:SM} shows our choices for four example models. The phase transitions are only weak to moderately strong which implies a speed of sound that is close to the value in a relativistic plasma.
\begin{table}[h]
	\begin{center}
		\begin{tabular}{ | c  || c | c | c | c || c | c |}
			\hline
			Model & $3\lambda/a_+$ & $E$ & $d$ & $T_n/T_{\rm cr}$ & $\aNEW_n$ & $c_{s,b}^2$ \\
			\hline
			\hline
			${\rm SM}_1$ & 10 & 0.3 & 0.2 & 0.9 & 0.0297 & 0.326 \\
			\hline
			${\rm SM}_2$ & 10 & 0.3 & 0.2 & 0.8 &  0.0498 & 0.331 \\
			\hline
			${\rm SM}_3$ & 3 & 0.3 & 0.2 & 0.9 & 0.00887 & 0.331 \\
			\hline
			${\rm SM}_4$ & 3 & 0.3 & 0.2 & 0.8 &  0.0149 & 0.333 \\
			\hline
		\end{tabular}
		\caption{\small\label{tab:SM} Parameters of the SM-like models. By construction the sound speed in the symmetric phase equals $c_{s,s}^2=1/3$.}
	\end{center}
\end{table}

The second model we study is a simplified version of a model with a two-step phase transition~\cite{Espinosa:2011ax} (see Section~\ref{sec:SMsinglet} for a more careful treatment for a two-step electroweak phase transition). The model has two scalar fields that break for example the electroweak symmetry and a $\mathbb{Z}_2$ symmetry. Although some symmetry is broken in both phases, we will still denote the phase that the field tunnels through first as `symmetric' and the second phase as `broken'. This time we neglect the cubic term. The pressure in the two phases can then be brought to the form
\bea
p_s(T) &=& \frac13 a_+ T^4 + (b_+ - d_+ T^2)^2 - b_-^2 \, , \nonumber \\
p_b(T) &=& \frac13 a_+ T^4 + (b_- - d_- T^2)^2 - b_-^2 \, ,
\eea
where we have subtracted the same cosmological constant in both phases. We can express one of the parameters using the critical temperature via $b_- - b_+ = T_{\rm cr}^2(d_- - d_+)$. Again the model has four relevant parameters, for example: $b_-/(\sqrt{a_+}T^2_{\rm cr})$, $d_-/\sqrt{a_+}$, $d_+/\sqrt{a_+}$ and $T_n/T_{\rm cr}$. Notice that in this model, the speed of sound in the symmetric phase also deviates from $c_{s,s}^2 = 1/3$. 

Table~\ref{tab:2step} shows our choices for four example models. The phase transitions are moderately strong to strong and the speed of sound in the broken phase in some of the models is below $c_{s,b}^2<1/4$.
\begin{table}[t]
	\begin{center}
		\begin{tabular}{ | c  || c | c | c | c || c | c | c|}
			\hline
			Model & $b_-/(\sqrt{a_+}T^2_{\rm cr})$ & $d_-/\sqrt{a_+}$ & $d_+/\sqrt{a_+}$ & $T_n/T_{\rm cr}$ & $\aNEW$ & $c_{s,b}^2$ & $c_{s,s}^2$ \\
			\hline
			\hline
			${\rm 2Step}_1$ & $0.4/\sqrt{3}$ & $0.2/\sqrt{3}$ & $0.1/\sqrt{3}$ & 0.9 & 0.0156 & 0.311 & 0.325\\
			\hline
			${\rm 2Step}_2$ & $0.4/\sqrt{3}$ & $0.2/\sqrt{3}$ & $0.1/\sqrt{3}$ & 0.7  & 0.0704 & 0.297 & 0.320\\
			\hline
			${\rm 2Step}_3$ & $0.5/\sqrt{3}$ & $0.4/\sqrt{3}$ & $0.2/\sqrt{3}$ & 0.9  & 0.0254 & 0.282 & 0.317\\
			\hline
			${\rm 2Step}_4$ & $0.5/\sqrt{3}$ & $0.4/\sqrt{3}$ & $0.2/\sqrt{3}$ & 0.7 & 0.159 & 0.245 & 0.306 \\
			\hline
		\end{tabular}
		\caption{\small\label{tab:2step} Parameters of the models with two-step phase transition. }  
	\end{center}
\end{table}
Table~\ref{tab:final} shows the kinetic energy fraction for out example models, as computed in the different methods, listed in Table~\ref{tab:methodComparison}. The M$_1$ column shows the full numerical result for $K$ and the results of the other methods are given as deviations from the full result.

\begin{table}[h!]
	\begin{center}
		\vspace{0.1cm}
		Deflagrations ($\xi_w=0.4$)
		
		\vspace*{0.5 cm}
		\begin{tabular}{ | c  || c || c |c |c |c |c |}
			\hline
			Model/Method & M$_1$ & M$_2$ & M$_3$ & M$_4$ & M$_5$ & M$_6$\\  
			\hline
			SM$_1$  &0.00208  & 
			$< 0.1 \%$ & -0.2 \% & -1.6 \% & -88.3 \% & 587.2 \% \\
			\hline
			SM$_2$ & 0.00536  & 
			$<0.1$ \% & -0.3 \% & -2.7 \% & -65.6 \% & 300.0 \% \\
			\hline
			SM$_3$ & 0.00021  & 
			$<0.1$ \% & -0.1 \% & -0.5 \% & -89.1 \% & 718.2 \% \\
			\hline
			SM$_4$ & 0.00057  & 
			$<0.1$ \% & -0.1 \% & -0.9 \% & -67.3 \% & 374.7 \% \\
			\hline
			2Step$_1$  & 0.00060  & 
			$<0.1$ \% & -2.0 \% & 1.3 \% & -90.6 \% & 726.4 \% \\
			\hline
			2Step$_2$  & 0.00883  & 
			$<0.1$ \% & -6.6 \% & -1.5 \% & -57.5 \% & 241.4 \% \\
			\hline
			2Step$_3$  & 0.00137  & 
			$<0.1$ \% & -2.8 \% & 5.3 \% & -92.3 \% & 749.6 \%  \\
			\hline			
			2Step$_4$ & 0.0285  & 
			-0.4 \% & -10.7 \% & 0.9 \% & -60.0 \% & 235.4 \% \\
			\hline			
		\end{tabular}
		\vspace*{0.5cm}
		
		Hybrids ($\xi_w=0.6$)
		
		\vspace*{0.5 cm}
		\begin{tabular}{ | c  || c || c |c |c |c |c |}
			\hline
			Model/Method & M$_1$ & M$_2$ & M$_3$ & M$_4$ & M$_5$ & M$_6$\\  
			\hline
			SM$_1$ & 0.00783 & 
			0.3 \% & 3.0 \% & 1.5 \% & -77.3 \% & 315.4 \% \\ 
			\hline
			SM$_2$ & 0.0159  & 
			$<0.1$ \% & 0.5 \% & -1.9 \% & -53.4 \% & 177.8 \% \\
			\hline
			SM$_3$ & 0.00157  & 
			0.2 \% & 2.2 \% & 1.8 \% & -79.4 \% & 347.5 \% \\
			\hline
			SM$_4$ & 0.00322  & 
			0.1 \% & 0.5 \% & -0.3 \% & -54.2 \% & 201.4 \% \\
			\hline
			2Step$_1$  & 0.00295 & 
			-0.2 \% & 10.5 \% & 14.2 \% & -78.7 \% & 406.4 \% \\
			\hline
			2Step$_2$  & 0.0210  & 
			0.1 \% & 2.3 \% & 7.8 \% & -41.4 \% & 172.2 \% \\
			\hline
			2Step$_3$  & 0.00443  & 
			-2.1 \% & 29.9 \% & 40.7 \% & -77.6 \% & 533.7 \% \\
			\hline
			2Step$_4$  & 0.0449  & 
			-1.1 \% & 8.6 \% & 22.7 \% & -38.6 \% & 209.6 \% \\
			\hline
		\end{tabular}
		
		\vspace*{0.5cm}
		Detonations ($\xi_w=0.9$)
		
		\vspace*{0.5 cm}
		\begin{tabular}{ | c  || c || c |c |c |c |c |}
			\hline
			Model/Method & M$_1$ &M$_2$ & M$_3$ & M$_4$ & M$_5$ & M$_6$\\  
			\hline
			SM$_1$ & 0.00143 & 0.5 \% & 5.0 \% & 3.6 \% & -88.5 \% & 713.3 \%  \\
			\hline
			SM$_2$ & 0.00401 & 0.4 \% & 1.7 \% & -0.7 \% & -66.7 \% & 351.9 \% \\
			\hline
			SM$_3$ & 0.00014 & $<0.1$ \% & 1.4 \% & 0.9 \% & -89.2 \% & 779.4 \% \\
			\hline
			SM$_4$ & 0.00039 & $<0.1$ \% & 0.4 \% & -0.3 \% & -67.9 \% & 405.1 \% \\
			\hline
			2Step$_1$ & 0.00036 & -0.2 \% & 13.6 \% & 17.4 \% & -89.5 \% & 945.2 \% \\
			\hline
			2Step$_2$ & 0.00563 & -0.8 \% & 15.7 \% & 21.9 \% & -50.0 \% & 366.2 \% \\
			\hline
			2Step$_3$ & 0.00070 & -0.8 \% & 36.0 \% & 47.3 \% & -89.9 \% & 1235.3 \%\\
			\hline
			2Step$_4$ & 0.0158 & -3.5 \% & 40.1 \% & 58.3 \% & -41.8 \% & 485.2 \% \\
			\hline
		\end{tabular} 
			
		\caption{\small\label{tab:final} Value of the kinetic energy fraction for our benchmark toy models. M$_1$ is the full numerical result and the results of M$_2$--M$_6$ are given as an error with respect to M$_1$. The model parameters are given in Appendix \ref{appendix:benchmarks}. The wall velocities we used are $\xi_w=0.4$, $\xi_w=0.6$, $\xi_w=0.9$ for deflagrations, hybrids and detonations respectively. The results for detonations are in agreement with Ref.~\cite{Giese:2020rtr}}
	\end{center}
\end{table}
\clearpage

\section{Code snippet for the efficiency in the template model}\label{App:code}

In this section we present a snippet of Python code to produce $\kappa (\aNEW_n,c_{s,s},c_{s,b})$ in the template model.

The functions {\tt mu} and {\tt getwow} encode the (special relativistic) relative velocity and the ratio of the enthalpies across the bubble wall. The function {\tt getvm} returns the fluid velocity behind the wall, $v_-$, and the expansion mode (2=detonation, 1=hybrid, 0=deflagration). The function {\tt dfdv} encodes the differential equation solved in the shock/rarefaction wave and returns $(d\xi/dv, dw/dv)$. The {\tt getKandWow} returns the enthalpy-weighted kinetic energy in the shock/rarefaction wave and the ratio between the enthalpy density at the start of the shock/rarefaction compared to its end (for the shocks, the end is in the phase in front of the shock; for the rarefaction wave, the enthalpy density is normalized to $1$ behind the wall and has to be rescaled in the other part of the code). The function {\tt alN} returns $\aNEW_n$ in the nucleation phase (in front of the shock) for a given $\aNEW_+$ value at the wall. This relation is specific for the template model. The function {\tt getalNwow} returns $\aNEW_n$ in the nucleation phase and the ratio of the enthalpies for fixed boundary conditions at the wall, namely, the fluid velocities behind/in front of the wall, the wall velocity, and the two speeds of sound. Finally, the function {\tt kappaNuMuModel} puts all the pieces together. It first determines the expansion mode and the fluid velocity behind the wall. In case there is a shock present, it varies the fluid velocity $v_+$ in front of the wall to reproduce the correct $\aNEW_n$ in the nucleation phase. Notice that this procedure leads to a unique solution, unlike varying for example $\aNEW_+$ at the wall. Finally, the contributions from the rarefaction wave are calculated and added. We tested our code with Python version 2.7.17 and scipy version 0.19.1.

\begin{table}[h!]
\tt 
\begin{tabular}{|r|l|} 
\hline
1 & import numpy as np \\
2 & from scipy.integrate import odeint \\
3 & from scipy.integrate import simps \\
4 & \\
5 & def mu(a,b): \\ 
6 & \quad  return (a-b)/(1.-a*b) \\
7 &  \\
8 & def getwow(a,b): \\
9 & \quad  return a/(1.-a**2)/b*(1.-b**2) \\
10 &  \\
11 & def getvm(al,vw,cs2b): \\
12 & \quad  if vw**2<cs2b: \\
13 & \quad\quad    return (vw,0) \\
14 & \quad  cc = 1.-3.*al+vw**2*(1./cs2b+3.*al) \\
15 & \quad  disc = -4.*vw**2/cs2b+cc**2 \\
16 & \quad  if (disc<0.)|(cc<0.): \\
17 & \quad\quad    return (np.sqrt(cs2b), 1) \\
18 & \quad  return ((cc+np.sqrt(disc))/2.*cs2b/vw, 2) \\
19 &  \\
20 & def dfdv(xiw, v, cs2): \\
21 & \quad  xi, w = xiw \\
22 & \quad  dxidv = (mu(xi,v)**2/cs2-1.) \\
23 & \quad  dxidv *= (1.-v*xi)*xi/2./v/(1.-v**2) \\
24 & \quad  dwdv = (1.+1./cs2)*mu(xi,v)*w/(1.-v**2) \\ 
25 & \quad  return [dxidv,dwdv] \\
26 &  \\
27 & def getKandWow(vw,v0,cs2): \\
28 & \quad  if v0==0: \\
29 &  \quad\quad   return 0,1 \\
30 & \quad  n = 8*1024  \# change accuracy here \\
31 & \quad  vs = np.linspace(v0, 0, n)  \\
32 & \quad  sol = odeint(dfdv, [vw,1.], vs, args=(cs2,)) \\
33 & \quad  xis, wows = (sol[:,0],sol[:,1]) \\
34 & \quad  if mu(vw,v0)*vw<=cs2: \\
35 &  \quad\quad   ll=max(int(sum(np.heaviside(cs2-(mu(xis,vs)*xis),0.0))),1) \\
36 &  \quad\quad   vs = vs[:ll] \\
37 &  \quad\quad   xis = xis[:ll] \\
38 &  \quad\quad   wows = wows[:ll]/wows[ll-1]*getwow(xis[-1], mu(xis[-1],vs[-1])) \\
39 & \quad  Kint = simps(wows*(xis*vs)**2/(1.-vs**2), xis) \\
40 &  \quad return (Kint*4./vw**3, wows[0])  \\
\hline 
\end{tabular}
\end{table}
\begin{table}[h!]
\tt 
\begin{tabular}{|r|l|} 
\hline 
41 & def alN(al,wow,cs2b,cs2s): \\
42 & \quad  da = (1./cs2b - 1./cs2s)/(1./cs2s + 1.)/3.\\
43 & \quad  return (al+da)*wow -da\\
44 & \\
45 & def getalNwow(vp,vm,vw,cs2b,cs2s):\\
46 & \quad  Ksh,wow = getKandWow(vw,mu(vw,vp),cs2s) \\
47 & \quad  al = (vp/vm-1.)*(vp*vm/cs2b - 1.)/(1-vp**2)/3.\\
48 & \quad  return (alN(al,wow,cs2b,cs2s), wow) \\
49 & \\
50 & def kappaNuMuModel(cs2b,cs2s,al,vw):\\
51 & \quad  vm, mode = getvm(al,vw,cs2b)\\
52 & \quad  if mode<2:\\
53 & \quad\quad    almax,wow = getalNwow(0,vm,vw,cs2b,cs2s)\\
54 & \quad\quad    if almax<al:\\
55 & \quad\quad\quad      print ("alpha too large for shock")\\
56 & \quad\quad\quad     return 0;\\
57 & \quad\quad    vp = min(cs2s/vw,vw) \\
58 & \quad\quad    almin,wow = getalNwow(vp,vm,vw,cs2b,cs2s)\\
59 & \quad\quad    if almin>al:\\
60 &  \quad\quad\quad     print ("alpha too small for shock")\\
61 &  \quad\quad\quad     return 0;\\
62 &  \quad\quad   iv = [[vp,almin],[0,almax]]\\
63 &  \quad\quad   while (abs(iv[1][0]-iv[0][0])>1e-7):\\
64 &  \quad\quad\quad     vpm = (iv[1][0]+iv[0][0])/2.\\
65 &  \quad\quad\quad     alm = getalNwow(vpm,vm,vw,cs2b,cs2s)[0]\\
66 &  \quad\quad\quad     if alm>al:\\
67 &  \quad\quad\quad\quad       iv = [iv[0],[vpm,alm]]\\
68 &   \quad\quad\quad    else:\\
69 &   \quad\quad\quad\quad      iv = [[vpm,alm],iv[1]]\\
70 & \quad\quad    vp = (iv[1][0]+iv[0][0])/2.\\
71 & \quad\quad    Ksh,wow = getKandWow(vw,mu(vw,vp),cs2s)\\
72 & \quad  else:\\
73 & \quad\quad    Ksh,wow,vp = (0,1,vw)\\
74 & \quad  if mode>0:\\
75 &  \quad\quad   Krf,wow3 = getKandWow(vw,mu(vw,vm),cs2b)\\
76 &  \quad\quad   Krf*= -wow*getwow(vp,vm)\\
77 &  \quad else:\\
78 &  \quad\quad   Krf = 0\\
79 &  \quad return (Ksh + Krf)/al\qquad\qquad\qquad\qquad\qquad\qquad\qquad\qquad\qquad\qquad\qquad\qquad\\
\hline 
\end{tabular}
\end{table}

\clearpage
\bibliographystyle{JHEP}
\bibliography{BeyondBag_2}

\end{document}